\documentclass{iaus}
\usepackage{graphicx}

\newcommand{\mas}{\mathrm{mas \, yr}^{-1}}

\title[Formation of dwarf galaxies]
{Formation of dwarf galaxies\\ and small-scale problems of CDM}

\author[Oleg Y. Gnedin]{Oleg Y. Gnedin}

\affiliation{University of Michigan, Department of Astronomy,
             Ann Arbor, MI 48109-1042, USA\break
             ognedin@umich.edu}

\pubyear{2006}
\volume{235}
\pagerange{1--4}
\date{Oct 27, 2006}
\jname{Galaxy Evolution across the Hubble Time}
\editors{F. Combes \& J. Palous, eds.}
\begin{document}

\maketitle

\begin{abstract}
The concordance cosmological model based on cold dark matter makes
definitive predictions for the growth of galaxies in the Universe,
which are being actively studied using numerical simulations.  These
predictions appear to contradict the observations of dwarf galaxies.
Dwarf dark matter halos are more numerous and have steeper central
density profiles than the observed galaxies.  The first of these
small-scale problems, the "missing satellites problem", can be
resolved by accounting for the low efficiency of gas cooling and star
formation in dwarf halos.  A newly-discovered class of HyperVelocity
Stars will soon allow us to test another generic prediction of CDM
models, the triaxial shapes of dark matter halos.  Measuring the
proper motions of HVS will probe the gravitational potential out to
100 kpc and will constrain the axis ratios and the orientation of the
Galactic halo.  
\end{abstract}

\firstsection
\section{Is the ``missing satellites problem'' still a problem?}

Hierarchical Cold Dark Matter (CDM) models predict that Milky
Way-sized halos contain several hundred dense, low-mass dark matter
satellites, an order of magnitude more than the number of observed
satellite galaxies in the Local Group.  If the CDM paradigm is
correct, this prediction implies that the Milky Way and Andromeda are
filled with numerous dark halos.  This has been termed the ``missing
satellites problem'' (\cite{klypin_etal99, moore_etal99}).  Despite
the recent discoveries of faint companion galaxies by the Sloan
Digital Sky Survey that nearly doubled the number of known dwarf
spheroidals (\cite{belokurov_etal06}), a large discrepancy between the
predicted and observed numbers remains (see Figure 1).

\begin{figure}
\begin{center}
\includegraphics[height=3.2in]{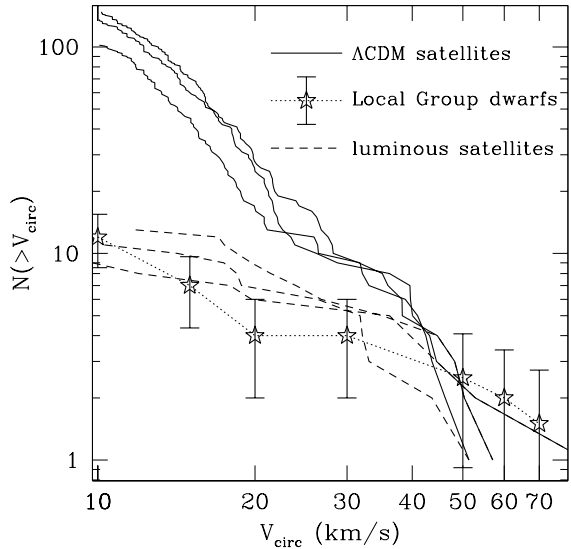}
  \caption{Cumulative velocity function of the dark matter satellites
  in three galactic halos ({\it solid lines\/}) compared to the average
  cumulative velocity function of dwarf galaxies around the Milky Way
  and Andromeda galaxies ({\it stars}).  Both observed and simulated
  objects are selected within the radius of $200h^{-1}\ \rm kpc$ from
  the center of their host. The dashed lines show the velocity
  function for the luminous satellites in our model.  The minimum
  stellar mass of the luminous satellites for the three hosts ranges
  from $\approx 10^5\ \rm M_{\odot}$ to $\approx 10^6\ \rm M_{\odot}$.
  From Kravtsov et al.\ (2004).}
  \label{fig:vf}
\end{center}
\end{figure}

In order to understand why most halos failed to become galaxies, we
need to understand their history.  In \cite{kravtsov_etal04}, we have
analyzed the dynamical evolution of the satellite halos in a
high-resolution cosmological simulation of three Milky Way-sized
halos.  We find that about 10\% of the substructure halos with the
present masses $\lesssim 10^8-10^9\rm\ M_{\odot}$ (circular velocities
$V_{\rm m}\lesssim 30\rm\ km/s)$ had considerably larger masses and
circular velocities when they formed at redshifts $z\gtrsim 2$.  After
the initial period of mass accretion while in isolation, these objects
experience dramatic mass loss due to tidal stripping.  Our analysis
shows that strong tidal interaction is often caused by actively
merging massive neighboring halos, even before the satellites are
accreted by their host halo.  These results indicate that some of the
systems that have small masses and circular velocities at $z=0$ could
have had masses comparable to those of the SMC and LMC in the
past. This can explain how the smallest dwarf spheroidal galaxies
observed in the Local Group were able to build up sizable stellar
masses in their seemingly shallow potential wells.

We have proposed a new model in which all of the luminous dwarf
galaxies in the Local Group are descendants of the relatively massive
($\gtrsim 10^9\rm \ M_{\odot}$) high-redshift systems, in which the
gas could cool efficiently by atomic line emission and which were not
significantly affected by the extragalactic ultraviolet radiation.  We
have constructed a semi-analytical galaxy formation model based on the
trajectories extracted from the simulation, which accounts for the
bursts of star formation after strong tidal shocks and the
inefficiency of gas cooling in halos with virial temperatures $T_{\rm
vir} \lesssim 10^4$~K.  Our model reproduces the abundance, spatial
distribution, and morphological segregation of the observed Milky Way
satellites, as well as their basic properties such as the stellar
masses and densities.

According to this model, all of the luminous satellites formed some
stars before the reionization ($z > 6$).  Afterwards, some galaxies
were massive enough to retain their gas and to keep forming stars
episodically, with new stars progressively more concentrated towards
the center.  These objects become dIrr and some dSph types.  The other
objects lost all of their gas during the reionization and contain only
the oldest stellar population, becoming fossil dSph galaxies
identified by \cite{gnedin_kravtsov06}.

\begin{figure}
\includegraphics[height=2.7in]{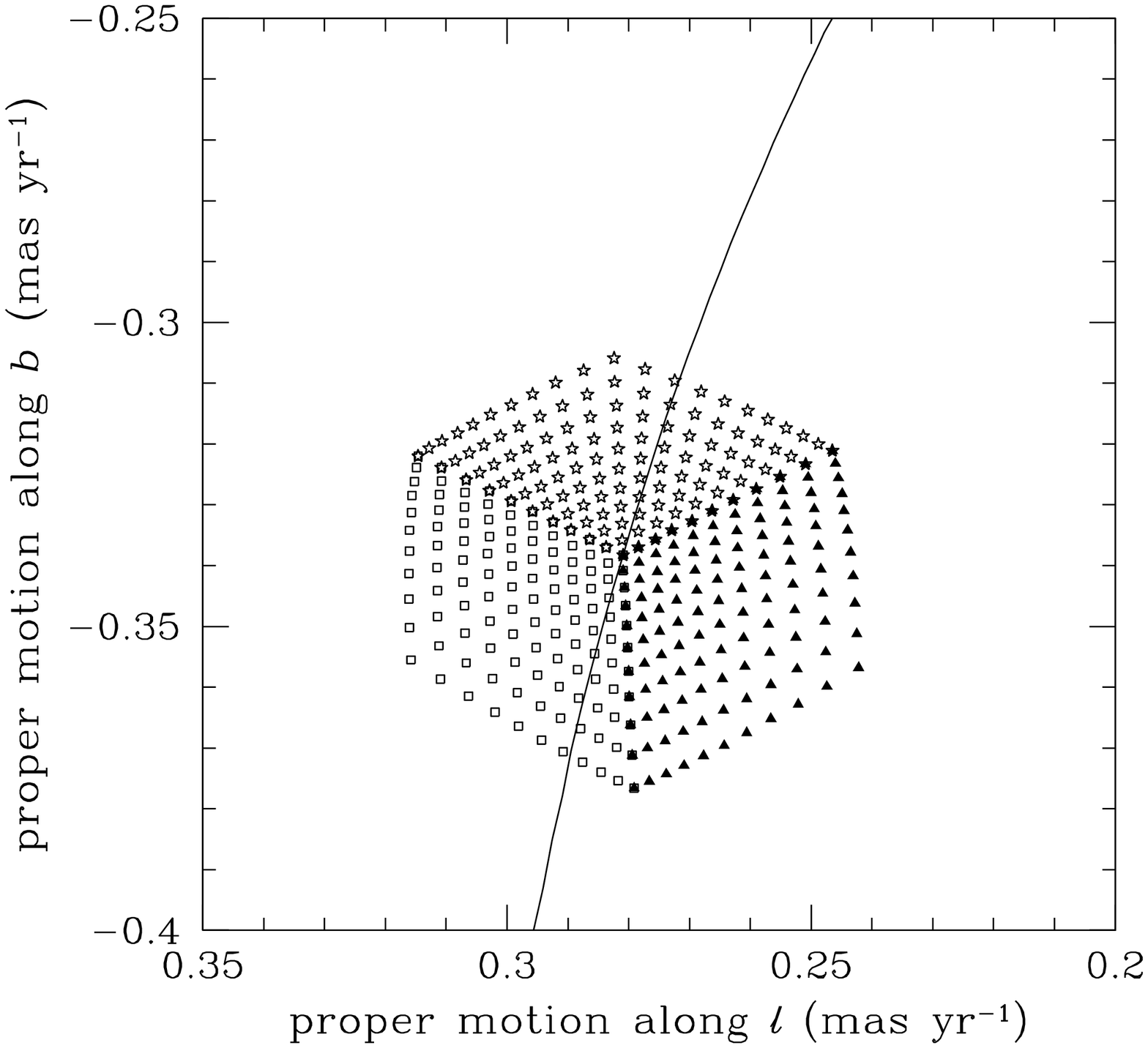}
\includegraphics[height=2.7in]{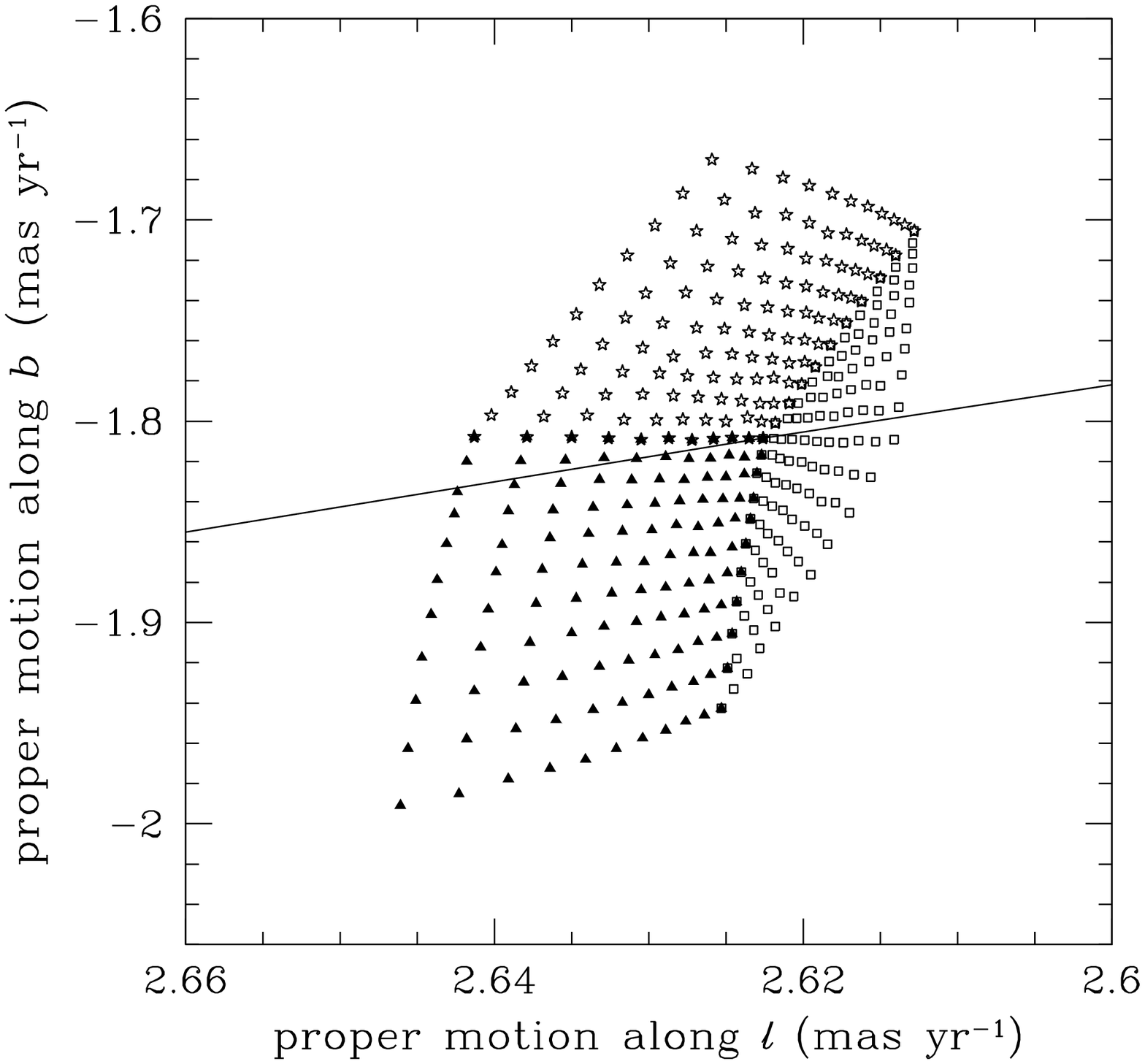}
\caption{Expected proper motions of HVS1 (left) and HVS2 (right) under
  a range of different assumptions about the shape and orientation of
  the Galactic dark-matter halo.  The two components of the proper
  motion are in the directions of Galactic coordinates, $l$ and $b$,
  and include the reflex proper motion due to the Sun's motion around
  the Galactic center.  The family of models with the halo major axis
  along the Galactic $X$-coordinate is shown by triangles, along the
  $Y$-coordinate by open squares, and along the $Z$-coordinate by open
  stars.  The solid line shows the predicted proper motions for the
  HVS1 distance from Earth varying from 61 (bottom) to 90~kpc (top),
  and the HVS2 distance from 18 to 20 kpc, assuming a spherical
  dark-matter halo.  Adapted from Gnedin et al.\ (2005).}
  \label{fig:hvs}
\end{figure}

\section{Probing the Galactic halo with HyperVelocity Stars}

A generic prediction of CDM models is that dark-matter halos are
triaxial, with density axis ratios in the range 0.5--0.8.  The only
currently known observational probe of the shape of the Galactic halo
is provided by tidal streams associated with satellite galaxies, such
as the Sagittarius dwarf spheroidal galaxy.  However, different
analyses of the Sagittarius stream data have produced conflicting
results, with claims that the Galactic halo is close to spherical
(\cite{johnston_etal05}), while others find that a minor-to-major axis
ratio as low as 0.6 cannot be ruled out for a prolate halo
(\cite{helmi04}).  Measurements of the halo shape close to the
Galactic disk are complicated by the dynamical effects of baryons,
tending to make even triaxial halos rounder at these radii
(\cite{kazantzidis_etal04}).  Another, independent measure of the halo
shape at larger distances is needed to test the nature of dark matter
and to confirm or falsify CDM models.

Such an independent test is now provided by the recently discovered
HyperVelocity Stars (\cite{brown_etal} 2005, 2006a, 2006b;
\cite{hirsch_etal05}; \cite{edelmann_etal05}).  The first three of
these extraordinary stars have heliocentric radial velocities above
$+700$ km s$^{-1}$ and were found in the course of spectroscopic
followups to all-sky surveys that have identified faint blue stars in
the Galactic halo.  The other four have velocities above $+540$ km
s$^{-1}$ and were discovered later in a targeted survey.  The
velocities of all HVSs exceed the plausible limit for a runaway star
ejected from a binary in which one component has undergone a supernova
explosion.  The only known mechanism for a star to obtain such an
extreme velocity is ejection from the deep potential of the
supermassive black hole at the Galactic center, as a result of
scattering with another star or tidal breakup of a binary
(\cite{hills88}).

Given its high-velocity ejection from the Galactic center, $\sim 1000$
km s$^{-1}$, the expected trajectory of a HVS in the Galaxy is nearly
a straight line.  However, the direction of the HVS present velocity
will deviate from being precisely radial due to departures from
spherical symmetry of the Galactic potential.  A precise measurement
of the three-dimensional motions of the HVSs thus probes the shape of
the Galactic halo mass distribution in a new way that is entirely
independent of any other technique attempted so far.

Figure \ref{fig:hvs} shows predictions for the proper motions of the
first two HVSs, consistent with their positions in the sky and the
observed line-of-sight velocities.  The orbits are calculated in a
generalized triaxial NFW potential, for different assumptions about
the density axis ratios.  For distant HVSs, the asymmetry of the
potential due to the flattened disk causes a smaller deflection than
that due to the triaxial halo.  The deflection contributed by the disk
peaks at $r \approx 10$~kpc but quickly declines at larger distances
where the disk density vanishes and the direction of the orbit aligns
with the velocity vector.  Interactions with the Galactic bar or
molecular clouds in the disk are even less important because of the
shorter lever arm.  On the other hand, the deflection due to the
triaxial halo {\it continues to accumulate along the whole
trajectory}.  Hence, the proper motions are sensitive to the halo
triaxiality and relatively insensitive to any uncertainties in the
mass model of the baryons.

We have started an HST program to measure astrometric proper motions
for five HVSs.  The first-epoch images are taken in Cycle 15 and the
second-epoch images will be taken late in Cycle 17.  With an almost
3-year baseline, we will easily detect the proper motion of HVS2 and
very likely those of the other HVSs.

With an expected measurement accuracy of $\sigma_\mu \approx 0.2\,
\mas$ (over 3 years) we can place useful constraints on the
orientation of the Galactic halo.  In the case of HVS2, for example,
if the major axis of the halo is aligned with the direction to the
Galactic center ($l=0^\circ$), the predicted components $\mu_b$ all
lie below $\mu_b < -1.8\, \mas$.  If the major axis is aligned with
the disk rotation axis, $\mu_b > -1.8\, \mas$.  If the major axis is
in the direction of solar rotation, then the predicted $\mu_l$
component of the proper motion is well constrained to be $\mu_l = 2.62
\pm 0.01\, \mas$.  Note also, that the proper motions are sensitive to
the distance to the HVS, and therefore, we will obtain a distance
estimate independent from the photometry.

Future astrometric satellites that could reach accuracy of $\sigma_\mu
\sim 0.01\, \mas$ will allows us to determine the halo axis ratios.
With two or more HVSs we will be able to break the
triaxiality-distance degeneracy and to constrain the axis ratios to
better than 20\%.

\begin{acknowledgments}
\noindent
I acknowledge the support of the American Astronomical Society and the
National Science Foundation in the form of an International Travel
Grant, and the support of the IAU through grant 11812.
\end{acknowledgments}


\begin{thebibliography}{}

\bibitem[Belokurov et al. 2006]{belokurov_etal06}
  {Belokurov, V. et al.}
  2006, \textit{ApJ} submitted, astro-ph/0608448

\bibitem[Brown et al.]{brown_etal}
  {Brown, W.~R., Geller, M.~J., Kenyon, S.~J. \& Kurtz, M.~J.}
  2005, \textit{ApJ} 622, L33; 
  2006a, \textit{ApJ} 640, L35;
  2006b, \textit{ApJ} 647, 303

\bibitem[Edelmann et al. 2005]{edelmann_etal05}
  {Edelmann, H. et al.} 
  2005, \textit{ApJ} 634, L181

\bibitem[Gnedin \& Kravtsov (2006)]{gnedin_kravtsov06}
  {Gnedin, N. Y. \& Kravtsov, A. V.}
  2006, \textit{ApJ} 645, 1054

\bibitem[Gnedin et al. (2005)]{gnedin_etal05}
  {Gnedin, O. Y., Gould, A., Miralda-Escude, J. \& Zentner, A. R.}
  2005, \textit{ApJ} 634, 344

\bibitem[Helmi 2004]{helmi04}
  {Helmi, A.}
  2004, \textit{MNRAS} 351, 643

\bibitem[Hills 1988]{hills88}
  {Hills, J. G.}
  1988, \textit{Nature} 331, 687

\bibitem[Hirsch et al. 2005]{hirsch_etal05}
  {Hirsch, H. A., Heber, U., O'Toole, S. J. \& Bresolin, F.}
  2005, \textit{A\&A} 444, L61

\bibitem[Johnston et al. 2005]{johnston_etal05}
  {Johnston, K. V., Law, D. R. \& Majewski, S. R.}
  2005, \textit{ApJ} 619, 800

\bibitem[Kazantzidis et al. 2004]{kazantzidis_etal04}
  {Kazantzidis, S., Kravtsov, A. V., Zentner, A. R., Allgood, B., Nagai, D.
   \& Moore, B.}
  2004, \textit{ApJ} 611, L73

\bibitem[Klypin et al. 1999]{klypin_etal99}
  {Klypin, A., Kravtsov, A.~V., Valenzuela, O. \& Prada, F.}
  1999, \textit{ApJ} 522, 82-92

\bibitem[Kravtsov, Gnedin \& Klypin (2004)]{kravtsov_etal04}
  {Kravtsov, A.~V., Gnedin, O.~Y. \& Klypin, A.~A.}
  2004, \textit{ApJ} 609, 482

\bibitem[Moore et al. 1999]{moore_etal99}
  {Moore, B. et al.}
  1999, \textit{ApJ} 524, L19

\end{thebibliography}
\end{document}